\documentclass[aps,10pt,prl,twocolumn,groupedaddress,showpacs,showkeys,amsfonts]{revtex4-1}
\usepackage{amsfonts}
\usepackage{amsmath}
\usepackage{amssymb}
\usepackage{graphics,graphicx}
\graphicspath{{./}}
\usepackage{epsf}
\usepackage{subfigure}
\usepackage{hyperref}
\usepackage[all]{hypcap}
\usepackage{color}
\newcommand{\eq}[1]{Eq.~\eqref{#1}}

\newcommand{\fig}[1]{Fig.~\ref{#1}}

\newcommand{\be}{\begin{equation}}
\newcommand{\ee}{\end{equation}}
\newcommand{\bem}{\begin{multline}}
\newcommand{\bea}{\begin{align}}
\newcommand{\eea}{\end{align}}

\def\cro#1{\left[ #1 \right]}
\def\pare#1{\left( #1 \right)}

\newcommand{\apporsm}[1]{the Appendix}

\begin{document}

\title{Can a periodically driven particle resist laser cooling and noise?}

\author{A. Maitra$^1$}
\author{D. Leibfried$^2$}
\author{D. Ullmo$^1$}
\author{H. Landa$^{3}$}
\email{haggaila@gmail.com}
\affiliation{$^1$LPTMS, CNRS, Univ.~Paris-Sud, Universit\'e Paris-Saclay, 91405 Orsay, France
\\$^2$National Institute of Standards and Technology, 325 Broadway, Boulder, Colorado 80305, USA \\ $^3$Institut de Physique Th\'{e}orique, Universit\'{e} Paris-Saclay, CEA, CNRS, 91191 Gif-sur-Yvette, France}

\begin{abstract}
Studying a single atomic ion confined in a time-dependent periodic anharmonic potential, we find large amplitude trajectories stable for millions of oscillation periods in the presence of stochastic laser cooling. The competition between energy gain from the  time-dependent drive and damping leads to the stabilization of such stochastic limit cycles. Instead of converging to the global minimum of the averaged potential, the steady-state phase-space distribution develops multiple peaks in the regions of phase space where the frequency of the motion is close to a multiple of the periodic drive. Such distinct nonequilibrium behaviour can be observed in realistic radio-frequency traps with laser-cooled ions, suggesting that Paul traps offer a well-controlled test-bed for studying transport and dynamics of microscopically driven systems.
\end{abstract}


\maketitle

An atomic ion trapped in near-vacuum is a highly isolated system whose quantum motion can be controlled exquisitely \cite{RevModPhys.85.1103}.
Notwithstanding this, it can also be a system where chaos and randomness at the microscopic level give rise to intriguing classical states of motion. 

Paul traps for atomic ions are based on radio-frequency ($10-200\,{\rm MHz}$) time-dependent potentials \cite{paul1990electromagnetic}. {The time-dependent drive affects the dynamics qualitatively and the trapping is based on dynamical stabilization, akin to stabilization of an inverted pendulum \cite{kapitsa1964collected,landau1976mechanics,rahav2003effective}.
In general, even for motion of one particle in one spatial dimension, a time-dependent drive renders the Hamiltonian phase-space effectively three-dimensional (3D), counting the time (which can be treated as periodic), the space coordinate and the momentum. 
For anharmonic potentials, this results in complex phase-space structures.

Laser cooling is widely used in ion trapping \cite{wineland1979laser, javanainen1980,javanainen1980a,javanainen1981laser, stenholm1986semiclassical, cirac1994laser,PhysRevA.64.063407,  marciante2010, PhysRevA.96.012519,PhysRevLett.119.043001, janacek2018effect}. With the cooling beam turned on, {the ion may be expected to be damped} to the {minimum of the effective} potential or, in some cases \cite{rfcooling}, heat up or diffuse to a larger amplitude where it may escape from the trap.
 However, even if the ion is cooled by the laser, the nonequilibrium nature of the dynamics implies in general that the peaks of its spatial probability distribution may not coincide with the minima of the potential. Rather, the distribution may develop new maxima, and complex stochastic limit cycles and hysteretic behaviour may emerge \cite{PhysRevA.74.023409,Kaplan09,PhysRevA.82.061402}.

In this Letter, we show that the anharmonicity in a periodically driven Paul trap can capture an ion at a large amplitude motion, corresponding to a {stable} limit cycle with sizeable basins of attraction in phase-space, even in the presence of damping by laser cooling and {the} associated randomness. 
The time-dependence of the potential is a critical ingredient for such stochastic limit cycles since, in contrast to time-independent confinement, it prevents the damping from erasing the signatures of the Hamiltonian phase-space in the (quasi) stationary state. 
Instead, the stochastic dynamics retain some of the more complex structure of the underlying Hamiltonian phase-space, with multiple peaks of the probability distribution of the ion emerging away from the effective potential minimum. 
Although we focus on a trapped ion, the basic required ingredients (time-dependence, anharmonicity, and weak damping) are relevant to many dynamically driven systems, as further discussed below.

 \begin{figure}
\includegraphics[width=3.3in]{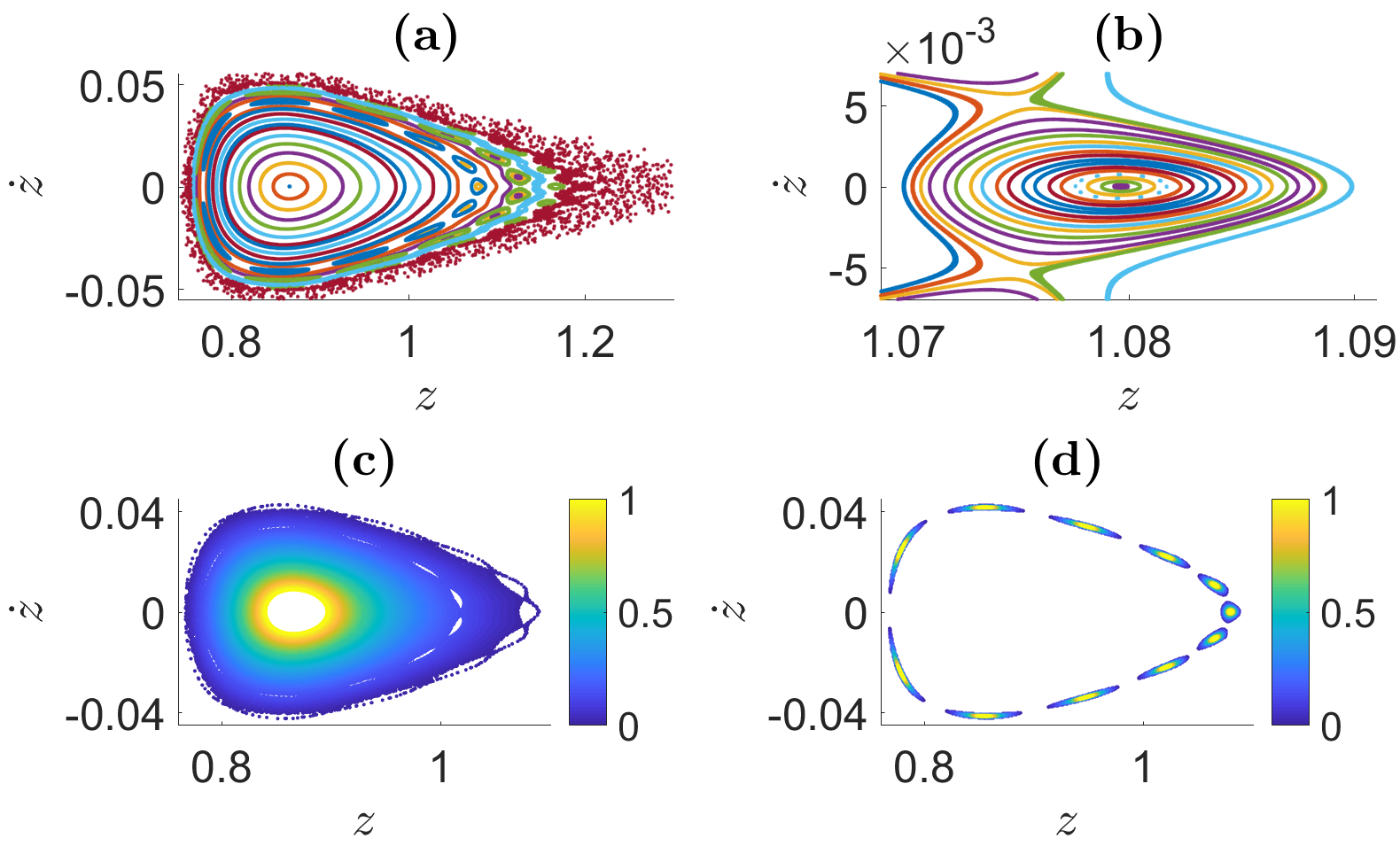}
\caption{Stroboscopic maps of the phase-space $\{z,\dot{z}\}$ in nondimensional units, obtained numerically by solving the equation of motion for the trap potential $V$ of \eq{eq:Vrf}, and plotting the points the trajectory goes through at times  $t=0\,{\rm mod}\,{\pi}$. (a) Mixed phase-space structure, wherein  the chaotic ``sea'', every closed curve and every ``island chain'' result from one initial condition. See the text for details of the parameters. (b) A close-up of the region around one island of the chain of $s=11$ islands.
 (c) With deterministic damping added [$\gamma=1\times 10^{-5}$ in \eq{Eq:eom}], an ion starting at the right-most curve in panel (b) will be slowly damped towards the effective potential minimum (the color code indicates the time along one slowly damping trajectory, 1 corresponds to $10^5$ drive periods). (d) An ion starting deeper within an island of a chain ($\{z=1.087,\dot{z}=0\}$), is captured in a large amplitude motion being damped towards the island centres [same color code as in (c)]. This motion remains stable when generalized to laser-cooling dynamics in three spatial dimensions, see \fig{fig:DopplerIslands}.
 }\label{fig:Islands}
\end{figure}

{\it Model.}
We consider dynamics in an effectively one-dimensional time-periodic potential in the presence of weak damping and first examine the stability of limit cycles based on an analytic expansion. We then use this expansion to numerically demonstrate that this mechanism remains robust under laser cooling in a realistic trap potential, when the remaining degrees of freedom corresponding to 3D confinement are taken into account.

In particular, we consider a time-dependent anharmonic potential,
\be V(z,t) = \frac{1}{2} a_z{\pare{z-z_s}}^2\\+ q V_2(z)\cos 2t,\label{eq:Vrf}\ee
{where} the time $t$ is in units of $2/\Omega$ {with} $\Omega$ {being} the angular frequency of the trap rf-drive and the coordinate $z$ is in units of a length-scale $d$, defined below. The charge and mass of the ion and the trap geometry parameters and voltages are absorbed into the nondimensional parameters $a_z$ and $q$.
Setting $z_s$ to be a point where $V_2(z)$ vanishes and choosing the parameters appropriately, makes $z_s$ a stationary point with stable motion in the phase-space around it. Since the periodic drive frequency has been rescaled to $2$, $V(z,t)$ is $\pi$-periodic.

In numerical simulations we take $V_2(z)=\frac{4}{\pi}\cro{\arctan\left(\frac{1}{2z}\right)-\arctan\left(\frac{3}{2z}\right)}$ to be {the potential} of a model surface trap \cite{rfchaos} along an axis perpendicular to the electrode plane (defined by z=0, with $d$ the width of two electrodes carrying the rf-modulated voltage). $V_2$ vanishes at $z_s=\sqrt{3}/{2}\approx 0.866$. The features discussed below can be observed with general anharmonic potentials; expanding $V_2$ to order $z^4$ about $z_s$ is sufficient to obtain the specific limit cycle studied here, although the details would vary.
{We take the values of the dimensionless parameters to be $ a_z=-0.0008$ and $ q\approx 0.87$. This corresponds to} realistic trap parameters (see below). For these nondimensional parameters, the analysis of \eq{eq:Vrf} can be applied to any ion species within such a trap, and any value chosen for $\Omega$ and $d$. Figure \ref{fig:Islands}(a) shows a stroboscopic map of the 3D phase-space, which is obtained by simulating the Hamiltonian dynamics with different initial conditions and stroboscopically plotting $\{z,\dot{z}\}$ at  times $t=0\,{\rm mod}\,{\pi}$.

The phase space is mixed with chaotic and integrable regions. Beyond the central region of closed curves that correspond to regular motion, there is a large chaotic ``sea'' and some sizeable ``island chains'' of regular motion forming about points where the anharmonic oscillation frequency of the ion {is} at a rational ratio with the drive frequency. Each closed curve is characterized by the existence of an action that is conserved during the Hamiltonian motion (in contrast, the energy is not conserved, since the potential is time-dependent), which is proportional to the enclosed area. 
Figure \ref{fig:Islands}(b) shows a close-up on the trajectories around one island in a chain which has $s=11$ islands [see \fig{fig:Islands}(a) and (d)]. Due to the chaotic regions, the phase-space is fragmented and the action cannot be defined globally, but it is well defined locally on the islands of a chain.

{\it Limit cycles.}
A trajectory {starting} at the center of one island of  a chain constitutes a periodic orbit repeating itself after $t=s\pi$, with the ion moving between the $s$ island centres in a fixed order.
When adding dissipation, the ion is attracted towards the periodic orbit from almost the entire island (defined as the largest area enclosed by a closed curve), a phenomenon previously studied mostly in terms of chaotic maps \cite{PhysRevE.54.71,lichtenberg2010regular}. 
To model this mechanism of trapping in the island chain, driven solely by damping, we add a friction coefficient $\gamma>0$ to the ion's equation of motion; 
\be \ddot{z}=F(z,t)-\gamma \dot{z},\qquad
F(z,t)=-\partial V(z,t)/\partial z.\label{Eq:eom}\ee
A numerical simulation of the time evolution using two (different) initial conditions is shown in \fig{fig:Islands}(c)-(d). In (c), the ion starts at $\{z=1.09,\dot{z}=0\}$, too far away from any island center to be attracted and is damped toward $z_s$, while in (d) it starts at $\{z=1.087,\dot{z}=0\}$, inside the closed trajectories around one of the island centres and gets trapped in the chain of $s=11$ islands.
Such limit cycles are a generic feature that results from the interplay of the nonlinearity of the Hamiltonian forces acting on the ion in the vicinity  of island chains, where the time-dependent drive counteracts the damping.

We now consider the dynamics of linearized perturbations about the limit cycle. Assume that  $\bar{z}(t)$ is an $s\pi$-periodic orbit that connects the island centres for $\gamma=0$, i.e.~$\ddot{\bar{z}}(t)=F(\bar{z}(t),t)$. We write
\be z(t)=\bar{z}(t)+u(t),\quad \bar{z}=\sum_n B_{2n}e^{i2nt/s},\label{eq:zt}\ee
 where $n\in\mathbb{Z}$ and $B_{2n}=B_{-2n}$ (time-reversal invariance is assumed since it simplifies the calculation). 
Substituting \eq{eq:zt} into \eq{Eq:eom} and linearizing the motion around $\bar{z}(t)$, we get
\be \ddot{u}+f(t) u=-\gamma\left( \dot{\bar{z}}(t)+\dot{u} \right),\quad f(t)\equiv -\frac{\partial F}{\partial z}\left( \bar{z}(t),t\right).\label{Eq:eomugamma}\ee
Substituting $u(t)$ by the ansatz $u(t)=w(t)e^{-\gamma t/2}$ gives
\be \ddot{w}+\left[f(t)-\frac{\gamma^2}{4}\right]w=g(t),\qquad g(t)\equiv-\gamma\dot{\bar{z}}(t)e^{\gamma t/2}.\label{Eq:eomw}\ee
The general solution for $w$ is composed of the sum of a particular solution growing exponentially as $e^{\gamma t/2}$, and the two linearly independent solutions of the homogeneous equation [with $g(t)=0$, see below].
Since $\bar{z}$ is $s\pi$-periodic and $F$ is $\pi$-periodic, the function $f(t)$  in \eq{Eq:eomugamma} can be Fourier expanded in the form $f=\sum F_{2n}e^{i2nt/s}$.
A particular solution of \eq{Eq:eomw} can be obtained by substituting 
$ w_0(t) = e^{\gamma t/2}\sum W_{2n}e^{i2nt/s}$,
 which gives an inhomogeneous system of recursion relations for $W_{2n}$, with a unique  solution under general conditions \cite{rfmodes}. The homogeneous equation in $w$ [\eq{Eq:eomw}] with $g(t)=0$, whose coefficients are periodic in time, is a Hill equation \cite{McLachlan}. The homogeneous solutions $w(t)$ determine the stability of $u(t)$ [since the exponential growth of the inhomogeneous $w_0$ is cancelled when going back to $u$]. In fact,  \eq{Eq:eomw} with $\gamma=g(t)=0$ determines the linear stability of the periodic orbit in the Hamiltonian case. If the area of the islands about the periodic orbit is not too small, perturbations about the periodic orbit are stable for a range of amplitudes of the motion, which implies that the motion would be linearly stable also for a small nonvanishing value of $\gamma$ in \eq{Eq:eomw}.
As \eq{Eq:eomw} shows, to leading order the damping is eliminated in the dynamics, and indeed we observe numerically that the ion is damped more slowly towards the island centres [\fig{fig:Islands}(d)], compared to
outside the island chain [\fig{fig:Islands}(c)].
  


{\it Laser-cooling}. To incorporate laser cooling more realistically, we assume a beam that is uniform over all positions in $\bar{z}$ and apply a recently developed semiclassical theory of laser-cooling that is valid for motion within the time-dependent potential of Paul traps \cite{rfcooling}. 
We approximate the ion as a two-level system, whose excited level has a decay rate (linewidth) $\Gamma$. The validity of our approach requires a low saturation of the internal transition, guaranteed by setting the saturation parameter $s_{\rm L}$ which is proportional to the laser intensity to $s_{\rm L}\ll 1$ \cite{rfcooling}. A saturation $s_{\rm L}\sim 0.1$  is often chosen in experiments, since it leads to the lowest final temperatures.
The instantaneous Doppler shift due to the ion's velocity determines its probability to absorb a photon at any phase-space point along the trajectory. For optical photons, many scattering events are required in order to change the  Hamiltonian action significantly.
 This allows us to describe the scattering as an adiabatic perturbations leading to a drift and diffusion of the ion between the tori (surfaces of constant action) of the Hamiltonian motion.

With a negative detuning $\Delta$ of the laser frequency from the internal electronic transition with a modulus that is large compared to $\Gamma$, an ion is efficiently cooled from in the region of high amplitude motion in the approximately integrable part of phase-space in a surface-electrode trap \cite{rfcooling}, whereas a laser detuned optimally for reaching the lowest temperatures (with $\Delta\sim -\Gamma/2$) could heat the ion out of the trap from this region. 
We numerically locate parameters for which a laser beam with $-\Delta\gg\Gamma$ can also capture the ion on trajectories within the island chain in the direct neighbourhood of a periodic orbit.

The coordinate $u(t)$ of \eq{eq:zt} represents linearized perturbations (a periodically driven harmonic oscillator) expanded about periodic orbit $\bar{z}(t)$, that can fluctuate, heat up and be damped by the laser. We numerically calculate $\bar{z}(t)$, $\dot{\bar{z}}(t)$  and $f(t)$ from \eq{eq:Vrf} [with $\gamma=0$], to obtain the Fourier expansion \cite{rfmodes} of the Hill oscillator of \eq{Eq:eomugamma}. This is a key step allowing us to introduce a canonical time-dependent transformation to the Hamiltonian action-angle coordinates $(I,\theta)$ describing the linearized motion about the periodic orbit. By the linearity of the expanded motion, the transformation can be obtained in analytic closed form, with the coefficients of the Fourier expansion calculated numerically \cite{rfions,rfmodes,zigzagexperiment}.
Averaging over the angle $\theta$, we obtain an effective Fokker-Planck equation for the probability distribution $P(I,t)$, which is a probability density function that depends on time and action only \cite{rfcooling},
\begin{align}
\frac{\partial}{\partial t} {P}(I,t) = -\frac{\partial}{\partial I} {S}(I,t)\equiv -\frac{\partial}{\partial I} \cro{\Pi_I {P}} +\frac{1}{2} \frac{\partial^2}{\partial I^2} \cro{\Pi_{II}{P}},
\label{eq:FP2}
\end{align}
with $S(I,t)$ a probability flux, $\Pi_I(I)$ an action drift coefficient and $\Pi_{II}(I)$ a diffusion coefficient. The calculation of $\Pi_I$ and $\Pi_{II}$ proceeds in a straightforward by using the formulas derived for a linear Floquet system in \cite{rfdist}.
 If we find a region of action where the ion remains bounded for a very long time (as determined by the Fokker-Planck dynamics), we can assume an approximately stationary probability distribution in the action, which then takes the form
\be P(I)\propto\left[\Pi_{II}(I)\right]^{-1}\exp\left\{2\int_0^I dI'{\Pi_I(I')}/{\Pi_{II}(I')} \right\}.\label{Eq:PofI}\ee

Taking concrete physical parameters we consider  a $^{24}{\rm Mg}^+$ ion. The nondimensional parameters  are (see \cite{rfchaos}) $a_z={4  e U_{\rm{DC}}}/{m  \Omega^2 {c_z}}$ and $ q={2eU_{\rm{rf}}}/{m d^2 \Omega^2}$, with $m$, $e$ the ion's mass and charge, $ U_{\rm{DC}}$, $U_{\rm{rf}}$ the voltages on the static and rf-modulated electrodes respectively, $c_z$ a geometric constant, and we set $ d=150\,{\rm \mu m}$ and  $\Omega=2\pi\times 20\,{\rm MHz}$. 
The resulting oscillation frequency at the effective minimum is $\omega_z\approx 2\pi\times 2.3\,{\rm MHz}$. The laser parameters used in our numerical simulation are ${k}= \cos(\phi)\times 2\pi / 280\,{\rm nm}^{-1}$ with $\phi\approx 0.4\pi$ giving the angle between the laser wavevector and the $z$ axis, and ${\Gamma}\approx 263\times 10^6\,{\rm s^{-1}}$. 


 \begin{figure}[t!]
\includegraphics[width=3.3in]{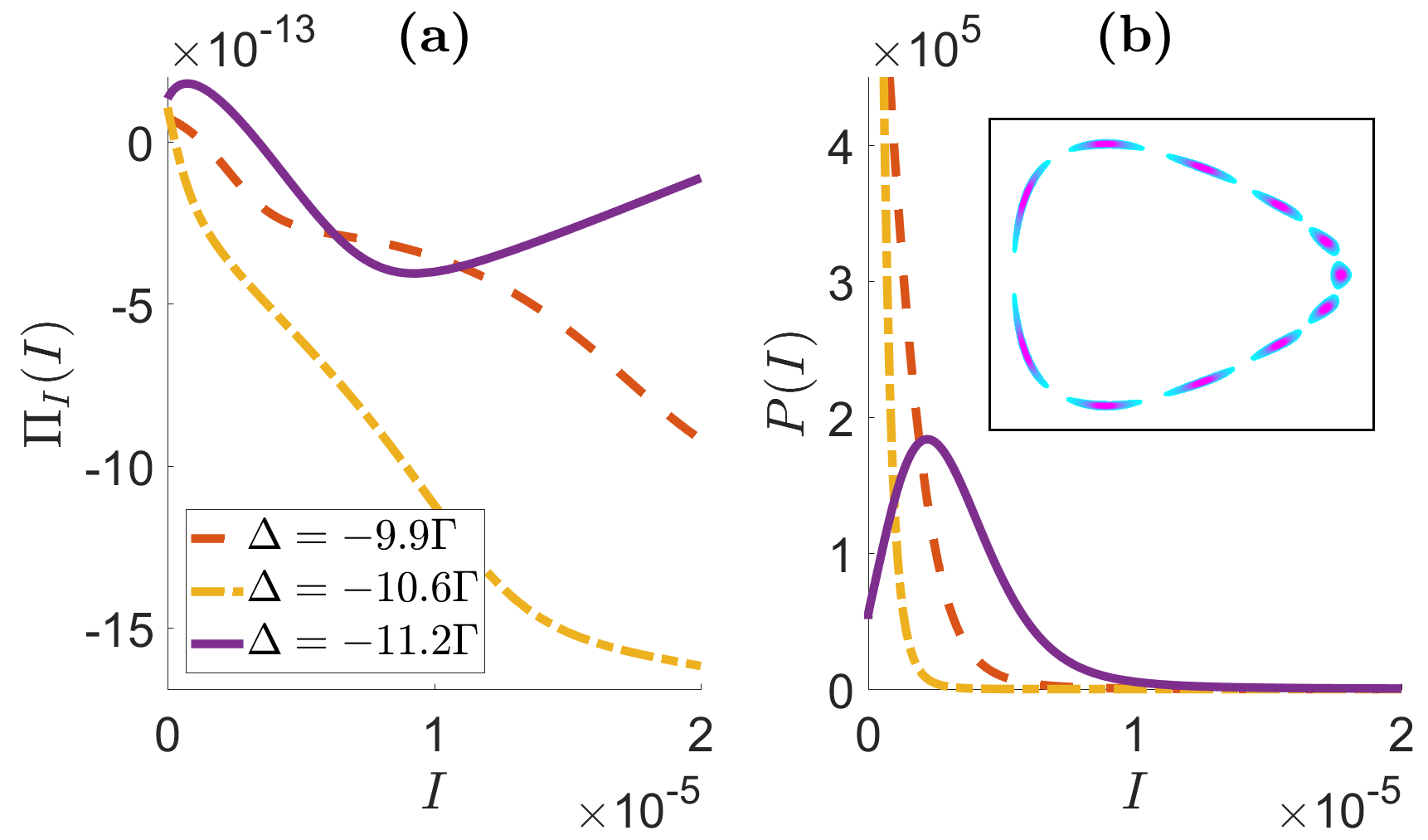}
\caption{(a) The action drift coefficient (in nondimensional units, see the text for the parameters) as a function of the action $I$ of linearized motion expanded about the periodic orbit going through the island centres of the chain with $s=11$ islands in \fig{fig:Islands}(b). Three values of the detuning are shown, with the $I$ axis extending roughly up to the size of (each) island. At action values where the drift coefficient is positive the ion is effectively heated by scattering laser photons, while it is being cooled where the drift coefficient is negative.
(b) The resulting approximately stationary probability distribution [\eq{Eq:PofI}]. For $-11\Gamma\lesssim\Delta\lesssim -9.8\Gamma$ the ion is well cooled within the island chain, with an exponential decay away from the center implying a lifetime estimated to be (for $\Delta = -10.6\Gamma$) of order tens of seconds. The inset shows a schematic depiction of the stroboscopic phase-space distribution (in $z$ and $\dot{z}$, at $t=0\,{\rm mod}\,{\pi}$).
}
\label{fig:DopplerIslands}
\end{figure}

We find a few ranges of a relatively large detuning $-\Delta\gg\Gamma$ for which the periodic orbit is stabilized. For $-11\Gamma\lesssim\Delta\lesssim -9.8\Gamma$, we find a stationary distribution in the action within the island chain with a strong (exponential) suppression of the probability of $I$ values away from the island centres. Figure \ref{fig:DopplerIslands}(a) presents 
the action drift coefficient for a few representative values of the detuning, showing that the action drift coefficient is negative throughout most of the island chain, indicating that the ion will drift towards the maximum of the quasi-stationary action distribution from any close by point on the island. Figure \ref{fig:DopplerIslands}(b) shows the corresponding approximately stationary distributions in action. Since this action is expanded about the island centres, and the angles are averaged over by assuming a uniform angle distribution, the corresponding (stroboscopic) phase-space distribution is that depicted schematically in the inset. For $\Delta=-11.2\Gamma$ we see that the probability density is peaked away from the island centre. In this case the exponential suppression of $P(I)$  is weaker than for the other detunings considered here, implying that the lifetime of the limit cycle would be possibly too short to be observed. Due to the stochastic nature of laser cooling, even an ion starting from outside the island chain has a finite probability to diffuse or drift onto one of the islands, however elucidating this mechanism is beyond the scope of the current work.

 Regarding the experimental observability of such a limit cycle, the lifetime within the island can be obtained by calculating the mean first passage time at the island boundary \cite{kampen2007}, and exceeds tens of seconds around $\Delta=-10.6\Gamma$ (we also find other ranges of the detuning with shorter lifetimes). 
 The mean photon scattering rate is $1.2\times 10^{6}\,{\rm s^{-1}}\times s_{\rm L}$, with $s_{\rm L}$ the saturation parameter. This rate can be contrasted with the rate at the effective trap minimum \cite{rfdist}, $(\Gamma s_{\rm L}/2)/\left(1+[2\Delta/\Gamma]^2\right)$, which gives $ 66\times 10^{6}\,{\rm s^{-1}}\times s_{\rm L}$ for the optimal detuning $\Delta=-\Gamma/2$, but $0.3\times 10^{6}\,{\rm s^{-1}}\times s_{\rm L}$ for $\Delta=-10.6\Gamma$. The photons can be detected and a Fourier transform of the fluoresence would show peaks at the $s$-subharmonics of the rf drive frequency. We have also simulated the Hamiltonian motion within the full time-dependent potential of the five-wire trap in three spatial dimensions for the parameters analyzed above, verifying that this mechanism is robust with large amplitudes of motion in the remaining coordinates. Island chains frequently develop in surface traps with voltages in the upper range of experimentally relevant values, and they can have very large relative sizes \cite{rfchaos}. 
 This is true also for other trap types, since the potential often attains some anharmonic contributions which become relevant above some energy or spatial scale. Hence we expect that such limit cycles can be observed in many existing traps and speculate that they may correspond to metastable extended spatial orbits that have been observed in Paul traps. 
 

 
To conclude, we have analyzed the dynamics of an ion driven by a periodic potential and stochastically kicked by a velocity-dependent force originating from photon scattering. The balance between the drive, the damping and the diffusion results in the ion stabilizing in a long lived limit cycle that corresponds to an extended orbit in space. The probability distribution develops new peaks within the islands of the mixed phase-space, centred about a periodic orbit frequency-locked to a rational multiple of the driving frequency. 
This provides an example of a system driven far-out-of-equilibrium that is nevertheless described by effective dynamics (coarse-grained over the angle), which approximately obey detailed-balance [since in steady state, the probability flux in action, $S(I,t)$ of \eq{eq:FP2}, must vanish \cite{rfdist}, a constraint obeyed by \eq{Eq:PofI}].
Our calculation manifests a method for obtaining the steady-state distributions for nonequilibrium underdamped systems \cite{PhysRevE.92.012716,fodor2016far, dotsenko2013two,grosberg2015nonequilibrium,
weber2016binary, mancois2018two,shankar2018hidden} through a mapping of the angle-averaged dynamics to that of an equilibrium system.

The dynamical mechanism we explored is formally close to models of Hamiltonian and Brownian ratchets \cite{denisov2014tunable}, which are basic models of transport \cite{reimann2002brownian, hanggi2009artificial}. 
Transport in a mixed phase-space is especially complex \cite{PhysRevLett.87.070601,PhysRevE.66.015207,PhysRevE.66.041104,PhysRevE.87.012918,PhysRevE.96.032204,tomkovivc2017experimental, firmbach20183d} and the ability to control and accurately measure motion in complex time-dependent potentials make ion-trap experiments suitable for quantitative tests of such ideas \cite{rfchaos}, extended even to many particles \cite{KinkTickle, liebchen2015interaction}. 
Our results can be generalized to frequency-locked limit-cycles in the relative coordinate of two interacting particles  \cite{Frequency_Locked_Orbits1993, Frequency_Locked_Orbits1994}, or in the rotation of a macroscopic particle, with $V_2$ a periodic function of the rotation angle \cite{PhysRevB.82.115441,delord2017strong}.

In the limit of small amplitudes, the coordinate $u(t)$ linearized about the island centres corresponds to a quantum parametric oscillator \cite{rfmodes,rfdist}. For the presented parameters, the mean action with $\Delta=-10.6\Gamma$ corresponds (by using the semiclassical relation $n\approx I/\hbar$, with $n$ the phonon number), to $\langle n \rangle\approx 170$ phonons, somewhat higher than the standard Doppler cooling limit $\langle n \rangle\approx 100$ of an identical ion whose secular frequency is similar, $2\pi\times 150\,{\rm kHz}$. Searching for similar limit cycles with larger driving frequencies and different parameters could facilitate cooling this motion to the quantum ground state using the well-developed tools of quantum optics in ion traps (e.g.~side-band cooling).  
This would open up a route for exploring quantum effects in phase-space \cite{bohigas1993,grossert2016experimental, tomkovivc2017experimental} with a single trapped ion beyond the limit of small-amplitude motion.




\begin{acknowledgments}
H.L. thanks Shmuel Fishman, Kirone Mallick and Roni Geffen for fruitful discussions, and acknowledges support by IRS-IQUPS of Universit\'{e} Paris-Saclay and by LabEx PALM under grant number ANR-10-LABX-0039-PALM.

\end{acknowledgments}

\bibliographystyle{hunsrt}
\bibliography{rf_cooling}

\end{document}